\documentclass[%
 reprint,
 amsmath,amssymb,
 aps,
prb,
floatfix,
]{revtex4-1}

\usepackage{graphicx}
\usepackage{dcolumn}
\usepackage{bm}

\begin{document}

\preprint{APS/123-QED}

\title{Sublattice Segregation of Hydrogen Adsorbates in Carbon Nanotubes}

\author{James Lawlor}
 \altaffiliation{School of Physics, Trinity College Dublin, Dublin 2, Ireland.}
 \email{jalawlor@tcd.ie}
\author{Mauro S. Ferreira}
\affiliation{
School of Physics, Trinity College Dublin, Dublin 2, Ireland.
}%
\affiliation{
CRANN, Trinity College Dublin, Dublin 2, Ireland.
}%

\date{\today}

\begin{abstract}

Recent experimental observations have reported that, rather than randomly distributed, nitrogen atoms may prefer to be located on one of the two sub-lattices of graphene. It has been suggested that such a preference may present a possible avenue to tailor the band gap of graphene whilst maintaining its excellent electronic transport properties. Among the proposed mechanisms to explain this effect is the suggestion that long-range inter-impurity interactions mediated by the conduction electrons of graphene may give rise to the asymmetry between sub-lattice occupations. Electron-mediated interactions are known to be prevalent not only between N atoms but also between impurities that are adsorbed to a specific location within the hexagonal structure, namely to the top of the carbon atom. Furthermore, this interaction is known to become more long-ranged as the dimensionality of the system is lowered. For that reason, in this paper we investigate whether a similar sub-lattice asymmetry appears in the case of metallic carbon nanotubes doped with Hydrogen adatoms. Our results indicate that similar sub-lattice asymmetries are observable and even more pronounced in small-diameter CNTs with a dilute concentration of adsorbates, diminishing with increasing diameter or impurity concentrations.


 \end{abstract}

\maketitle


\section{\label{sec:intro}Introduction}
 
 Since the discovery of Carbon Nanotubes (CNTs) in 1991 \cite{iijima1991helical} there has been a huge amount of interest in exploiting their unique mechanical and electronic properties, with a wide range of applications such as energy storage \cite{liu1999hydrogen,Ströbel2006781,yurum2009storage}
, composites \cite{de2013carbon,SMLL:SMLL201203252} and electronics \cite{cao2013single,franklin2013electronics,shulaker2013carbon}. 
 The extraordinary properties of CNTs come from their atomic structure, being constructed of carbon atoms bonded via sp$^2$ bonds and forming a tubular hexagonal lattice - a quasi 1D structure.  
 It is well known that the introduction of dopants to nanostructures can change their electronic properties, and current research with doping CNTs aims to tailor the electronic structure through controlling the dopant positioning \cite{khalfoun2014long, khalfoun2015transport}.
 \par
 Graphene, another novel carbon-based material, has had a similar level of research interest. 
 While it has the same hexagonal structure as CNTs, it is totally flat - indeed it can be imagined that CNTs are simply rolled up sheets of graphene. 
 One common feature is that both materials, due to their hexagonal lattice structure, can be modelled as having two inter-penetrating triangular sublattices, is illustrated in Fig. \ref{fig:schematic} .
 Recent experiments of nitrogen doping in chemical vapour deposition-grown graphene have shown that dopants have a preference for occupying the same sublattice, forming large domains of many dopants on only one sublattice \cite{zhao2011visualizing,lv2012nitrogen,zabet2014segregation, wang2012review}. 
 Even post-synthesis doping by direct ion implantation followed by heat treatment has shown preference for same sublattice configurations \cite{telychko2014achieving}, suggesting the phenomenon can occur in other scenarios.
 Indeed, observations of same sublattice preference have been reported for low concentration Molybdenum impurities \cite{wan2013incorporating} and high-coverage hydrogenated graphene \cite{lin2015direct}.
 Nevertheless there are many situations where no asymmetry effect is observed, and more work needs to be done to clarify the situations in which one would expect to find the sublattice asymmetry in doping and to what degree.
 
 Despite this uncertainty, these findings have been particularly interesting to the graphene community for many reasons.
 Firstly, it has been shown that this asymmetry in doping can be exploited in order to control the electronic properties of the material introducing the possibility to open a band gap whilst preserving the excellent transport properties of its pristine state \cite{lherbier2013electronic,lawlor2014sublatticereview,radchenko2014effects,botello2013modeling}.
 Other work has investigated the novel properties that arise from edges and grain boundaries in the sample \cite{aktor2015electronic} and its potential use in ferromagnetic response from doped graphene \cite{park2013spin}.
 Similarly, research with metallic CNTs doped with nitrogen has shown a large difference in transport response between segregated and unsegregated doping patterns \cite{khalfoun2014long}.
 The question of whether it is experimentally feasible to induce a ferromagnetic state in either doped CNTs \cite{pei2006effects,lu2011adsorptions,santos2011magnetism} and graphene \cite{lieb2004two,palacios2008vacancy,kumazaki2008tight}
 hydrogen \cite{casolo2009understanding}, which arises from sublattice segregation of defects, can also be re-examined.
 
 The mechanism behind the segregation effect is debated and several theories have been proposed, for example the effect could occur through nucleation during the CVD growth process \cite{zabet2014segregation,deretzis2014origin} or alternatively from impurity interactions mediated by electrons in the host \cite{lawlor2013friedel,lawlor2014sublattice} where it is known these interactions can be long-ranged \cite{lambin2012long} like the related Ruderman-Kittel-Kasuya-Yosida (RKKY) interaction for magnetic impurities \cite{klinovaja2013rkky,liu2011fe,sherafati2011analytical,bacsi2010local,kopylov2011transport}.
 The latter explanation captures the experimental observation of nitrogen asymmetry in both CVD synthesis and ion implantation with subsequent heat treatment, however the two mechanisms may be complimentary to each other.
 \par
 
 We propose that if the second theory is correct that a similar effect should not only be observable in CNTs, but be more robust. 
 Hypothetically the effect could occur in many metallic nanostructures but the mathematical formalism of the interaction and its similarity to the RKKY interaction suggests the effect should be more pronounced in CNTs.
 The RKKY interaction is an effect by which two magnetic impurities can have their moments aligned or anti-aligned depending on their separation \cite{ruderman1954indirect,kasuya1956electrical,yosida1957magnetic}, and this effect is mediated by conduction electrons in the host.
 Additionally, the decay of the RKKY between two impurities separated by a distance (D) in any system follows its dimensionality (d), generally going as the function $D^{-d}$, 
 hence in CNTs the decay is $D^{-1}$ compared to a quicker $D^{-2}$ in graphene \cite{power2013indirect,duffy2014variable,PhysRevB.36.3948,costa2005indirect}.
 As both the RKKY and the proposed asymmetry effect arise from symmetry breaking in the lattice causing so-called Friedel oscillations in the local density of states \cite{lawlor2013friedel,bacsi2010local} it seems that CNTs are a natural candidate to test the latter effect. 
 \par
 
 In this work we will demonstrate that sublattice asymmetry of adsorbed atomic hydrogen atoms in CNTs should arise in a similar way to substitutional nitrogen dopants in graphene. Monoatomic hydrogen is perhaps the 'simplest' dopant to consider from both an experimental and a theoretical point of view, and the doping of CNTs and other nanostructures with hydrogen is well understood due to the potential use for future energy storage \cite{jones1997storage,liu1999hydrogen,orivnakova2011recent,dutta2014review,liu1999hydrogen,Ströbel2006781,yurum2009storage,bowman2002metallic,gross2002catalyzed,chen2013atomic,froudakis2011hydrogen,nikitin2005hydrogenation,schlapbach2001hydrogen,johns2012atomic,yang2002ab}. 
 Although nitrogen doped CNTs are well documented in the literature, hydrogen is an adsorbed impurity, instead of being an sp2 bonded substitutional impurity like nitrogen.
 The advantages of using hydrogen instead of nitrogen is that the doping can be applied post-synthesis of the graphene sheet and that the adsorbates can migrate atop the graphene with little energy \cite{zhang2007ab,vehvilainen2009multiscale,borodin2011hydrogen}. 
 This is not the case for nitrogen \cite{zabet2014segregation,tison2015electronic,willke2014short, joucken2012localized}, although there is limited evidence to suggest that a subtle version of the sublattice asymmetry effect can be produced using high temperature annealing after post synthesis doping  \cite{telychko2014achieving}.

 The mathematical framework developed in recent work for the asymmetry effect in graphene \cite{lawlor2014sublattice} will be used as a basis for Monte-Carlo simulations where the additional effects of the geometry and size of the CNTs can be investigated, along with the propensity for the adsorbed hydrogen to cluster.
  
\section{\label{sec:methods}Methods}
 
\subsection{\label{sec:lloyd} Tight-binding model and inter-impurity interactions}

\begin{figure}[ht]
  \centering
    \includegraphics[width=0.45\textwidth]{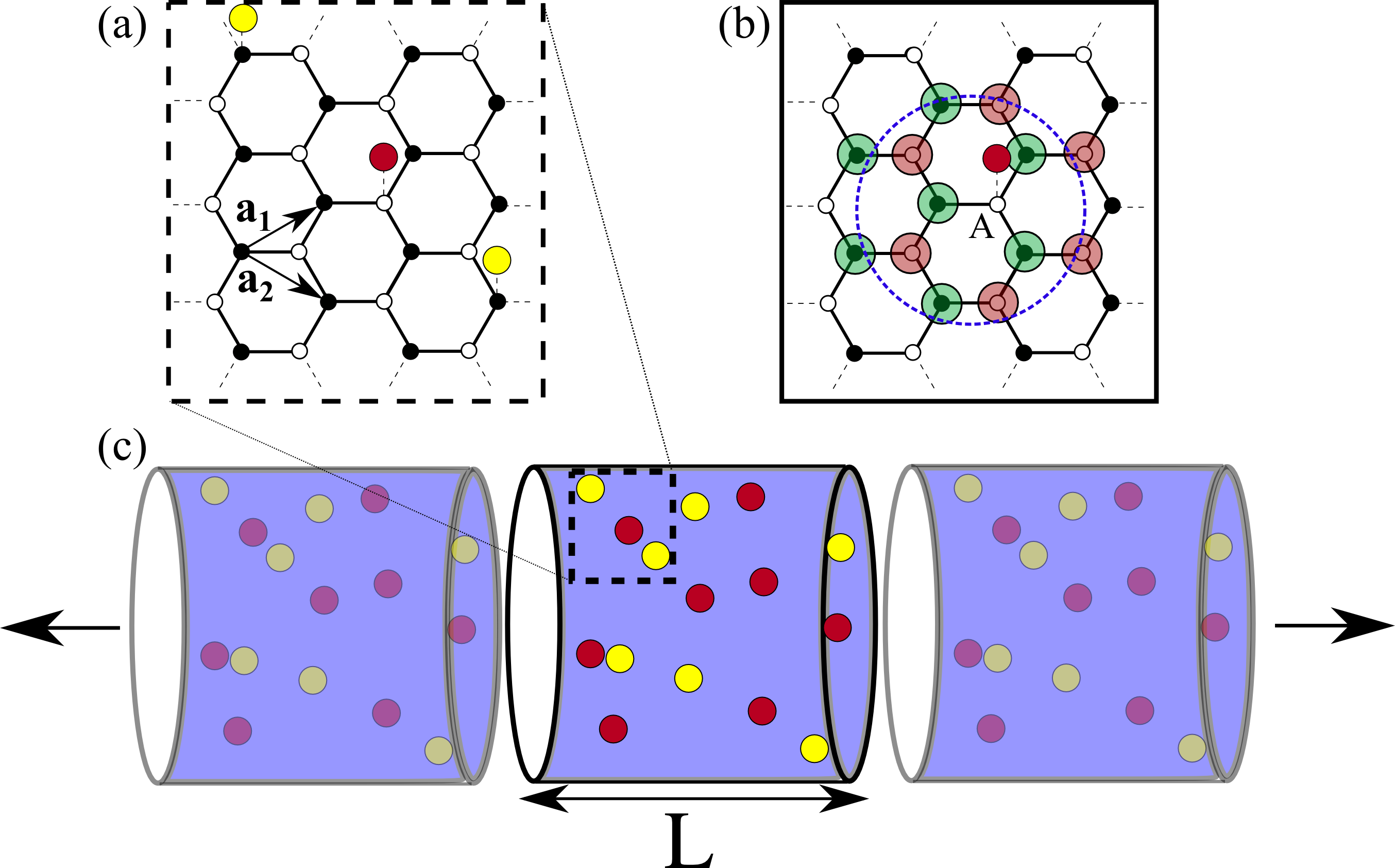}
  \caption{ Schematic representation of a nanotube wrapped in the zigzag direction (ZZNT) doped with atomic hydrogen impurities.
  (a) shows three adsorbates bonded to different carbon host sites and colour coded depending on the sublattice of the host, also shown are the lattice vectors $\mathbf{a_1}$ and $\mathbf{a_2}$.
  This is a close-up of the dashed box region in the larger ZZNT system shown in (c), where an ensemble of $N=15$ impurities are shown in a central region of length $L$. 
  Throughout the text we will refer to the degree of sublattice asymmetry in such a system using the percentage of impurities on the majority sublattice.
  It is evident in this example case that there are 8 impurities on the white sublattice and 7 on the black sublattice, so the measured asymmetry is $\frac{8}{7+8} = 53.33$\% and the doping appears random.
  The central region is replicated along the axial direction, indicated by the ghosted copies to the left and right, in order to minimise finite-size effects.
  To exaggerate the concept of the central and replicated cells, each one is shown as a separate cylinder however in the model no such physical separation exists. 
 The top-right schematic (b) shows the possible locations within the dashed blue circle region that an impurity at site A in the central unit cell can move to if randomly selected in the iterative Monte-Carlo procedure described in the main text, colour coded for opposite (green) and same (red) sublattice sites.
 }
  \label{fig:schematic}
\end{figure}

 While the mathematical framework for the asymmetry effect has been covered in-depth in recent work \cite{lawlor2013friedel,lawlor2014sublattice}, an overview is presented here for self-completeness. 
 Using a nearest-neighbour tight-binding method for graphene yields a Hamiltonian of the form $\hat{H} = - \sum_{ij} t_{ij}$ describing the hopping between nearest neighbour atoms $i$ and $j$ with the energy $t_{ij} = 2.7eV$.
 From this one can derive the Green Functions (GFs) for CNTs with both armchair (ACNT) and zigzag (ZZNT) geometries through inversion of the graphene Hamiltonian and enforcing certain periodic boundary conditions, a process explained in detail in the literature \cite{costa2005indirect,Lawlor201548}.
 As CNTs can vary in circumferential size, we follow the convention of ACNT$_n$ and ZZNT$_m$ to refer to armchair and zigzag tubes of circumferences defined by the lattice vectors $\mathbf{a_1}$ and $\mathbf{a_2}$ in Fig. \ref{fig:schematic} (a).
 If the circumference is defined as $C = r_1 \mathbf{a_1} + r_2 \mathbf{a_2}$ it follows that a ANCT$_m$ (ZZNT$_m$) has a circumference defined by $r_1 = m$, $r_2 = m$ ($r_1 = m$, $r_2 = -m$).
 
 Tight-binding is one of the more computationally viable methods for quantum mechanical Monte-Carlo simulations, which rely on a huge amount of random samples of a system to arrive at a reliable estimate of the ground state, due to its simplicity. 
 Additionally, it is well known that the electronic structures of graphene and CNTs are described particularly well by the tight-binding method \cite{reich2002tight}. 
 In the tight-binding regime the addition of impurities and their effect on a pristine system with GF $\hat{g}$ is a relatively straightforward process by use of the Dyson Equation and a suitable description of the impurities. 
 Atomic hydrogen adsorbates can be characterised thus by an onsite energy $\epsilon_a = 0.66t$ relative to the Dirac point energy of graphene and higher than the system Fermi energy, and hopping integral $\tau = 2.2t$ with a single carbon host site \cite{robinson2008adsorbate}. In the case of nitrogen it was found that the exact parameterisation of the impurity can affect the quantitative results but that the qualitative behaviour remains the same.
 The parametrisation of hydrogen used here captures the close range behaviour expected of pairs of hydrogens in graphene and CNTs whereby they prefer to occupy opposite sublattices to each other \cite{buchs2007creation,vehvilainen2009multiscale,zhang2007ab,pei2006effects,roman2007hydrogen}.
 The aim of this work is to model whether sublattice asymmetry can occur in nanotubes with adsorbed hydrogen impurities, so the behaviour produced by this parameterisation should be sufficient for this purpose.
 Whilst numerous calculations of pairs of hydrogen adatoms, based on density functional theory (DFT), have shown that opposite sublattice configurations are more preferable, these are inadequate for considering the long-range disorder to be modelled here.
 Firstly, the symmetry breaking caused by the adsorbtion of the dopant and the subsequent $E_F$ shift will induce Friedel oscillations in the electron density leading to a long-range spatial oscillation in the energetics of the preferred configuration, and that it is the long-range behaviour that leads to the asymmetry, as we will show.
 Also, the periodic nature of the DFT unit cell used to calculate these preferred configurations, effectively forming a superlattice, is also problematic when considering these interactions especially when a small cell is used.

 To calculate the most energetically favourable configuration for two or more impurities, the total energy change of the system must be calculated. 
 When impurities are added to a system this energy can be calculated through the Lloyd Formula method \cite{lloyd1967wave}, and has the general form

\begin{equation}
  \Delta E_N = \frac{2}{\pi} \int_{-\infty}^{E_F} dE \, {\rm Im} \ln \det \left(\hat{I} - \hat{g}(E) \hat{V} \right)
\label{eq:de}
\end{equation}
where $\hat{g}$ is the GF of the pristine system, $\hat{V}$ is a matrix describing the perturbation introducing $N$ impurities and $E_F$ is the Fermi energy.

 For a system with two impurities the total system energy change $\Delta E_2$ can be separated into a relative separation and configuration dependent term $\delta E$ arising from the interaction of the impurities, and a term $\Delta E_1$ which is the energetic cost of adding a single impurity in the system and is therefore position independent.
 As a result it can be shown that
 \begin{equation}
  \delta E = \Delta E_2 - 2 \Delta E_1 
\label{eq:deab}
\end{equation}
 which will henceforth be referred to as the "interaction energy" and has a long-range decaying oscillatory functional form \cite{lawlor2013friedel} $\delta E \sim \frac{\cos{2 Q_F D}}{D}$ where $Q_F$ is associated with the Fermi wavevector. 
 The calculation of the ground state for a system with only 2 impurities is relatively simple computationally, but the complexity of this calculation increases with $N!$ for N impurities making the exact calculation prohibitively expensive. 
 It is possible, however, to approximate the ground state using Monte-Carlo methods, and we demonstrate our methods for doing this in the following section.
 
\subsection{\label{sec:mcmethods}Monte-Carlo modelling of a finite concentration of impurities}
 The strength of Monte-Carlo methods are that they can be used to simulate large complex systems where arriving at an exact numerical solution is difficult due to the scale or complexity of the problem. These techniques have previously been used to look at sublattice segregation in general adsorbates on graphene \cite{cheianov2010sublattice}, and they will be used in this work as outlined below.
 
 Consider a section of nanotube of length $L$ with a distribution of $N$ adsorbates randomly attached to host carbon atoms in the system such as that in Fig. \ref{fig:schematic}. Using the expression for $\delta E$ and assuming only pairwise interactions, as higher order interactions will decay quicker, one can approximate the ground state by applying many iterations of choosing an impurity, summing all pairwise interaction energies between itself and the other impurities, and comparing this energy to the same energy when the impurity has been moved to a nearby site (hereafter referred to as $\delta E_\text{Switch}$), as demonstrated in Fig. \ref{fig:schematic} (b).
 If the second energy is lower then we assume that this lowers the total system energy, and the impurity is moved. If the first energy is lower then no change is made.
 This technique is similar to that used for the computational approach to the Ising model of ferromagnetism in a 2D lattice.
 The effect of a Boltzmann Temperature $T_B$ can also be introduced, whereby there is a finite probability $\sim e^{\delta E_\text{Switch} / k_B T_B}$ to move the impurity even though the first energy may be lower, where $k_B$ is the Boltzmann constant.
 It should be clarified that this iterative method of moving adsorbates is not the same as a dynamics calculation. While energy barriers for diffusion have been calculated to be on the order of $\leq 1$eV \cite{zhang2007ab, vehvilainen2009multiscale,borodin2011hydrogen}, we choose to ignore these as the aim of this work is to find the ground state of the system and not model the physical diffusion of adatoms. 
 Moreover, recent research has shown it is possible to make diffusion easier by doping CNTs with substitutional nitrogen \cite{zhang2007ab}.

 If a concentration of impurities ($\rho$) is added to the system it is natural to assume the number of electrons and therefore the Fermi Energy $E_F$ will change. 
 $E_F$ is adjusted with concentration according to DFT results for different length CNTs doped with a single hydrogen. 
 It is found that the relationship is linear and approximated well up to $\rho \sim 10\%$ by assuming that there is $\frac{1}{12}$ electron transfer to the host system per hydrogen and that the density of states profile is relatively unchanged.
Other methods of finding this $E_F$ shift have found a higher charge transfer from adatom to the graphene \cite{pike2014tight} which would result in shorter oscillation periods and a reduction of the predicted segregation effect.
 
 To minimize finite-size effects which will be intrinsic to our methodology, it is necessary to impose some extra conditions.
 Firstly the central region and the impurities are replicated along the axis of the nanotube once in each direction in order to approximate the effect of an infinite system.
 A schematic of this is shown in Fig. \ref{fig:schematic} (c).
 When iterating, if an impurity in the central region is moved then so are their 'twins' in the replicated sections.
 Because of the oscillatory nature of the interaction we impose a cut-off of one oscillation period of $\delta E$, beyond which the interaction is assumed to be zero. This is to ensure an even contribution to the interaction energy for all the impurities regardless of their position in the central region.
 Due to the scaling between length $L$, impurity concentration $\rho$ and Fermi Energy $E_F$, this cut-off length is approximately the same as $L$ and for the example system presented in Fig. \ref{fig:deab} is approximately $D = 180a$, where $a=0.14$nm is the carbon-carbon bond length of graphene, and this can be seen more clearly in Fig. \ref{fig:deab} (a). 
 Using the mathematical techniques developed in previous work concerning Friedel oscillations and the sublattice asymmetry effect in nitrogen-doped graphene \cite{lawlor2013friedel,lawlor2014sublattice} it can be shown that the oscillation period changes as $\frac{1}{\rho}$.

 The following calculations were done using approximately $N=40$ impurities for ACNTs and approximately $N=30$ for ZZNTs, $10 N^2$ iterations and 250 randomly generated systems for each concentration and temperature combination. 
 The cut-off, directly linked to $E_F$ and thus $\rho$, determines the exact value of $L$ which in turn affects the value of $N$, hence why there is a difference in $N$ between ACNT and ZZNT systems.

\section{\label{sec:results}Results}

\subsection{\label{sec:twoimps} Two hydrogen impurities}

\begin{figure}[ht]
  \centering
    \includegraphics[width=0.45\textwidth]{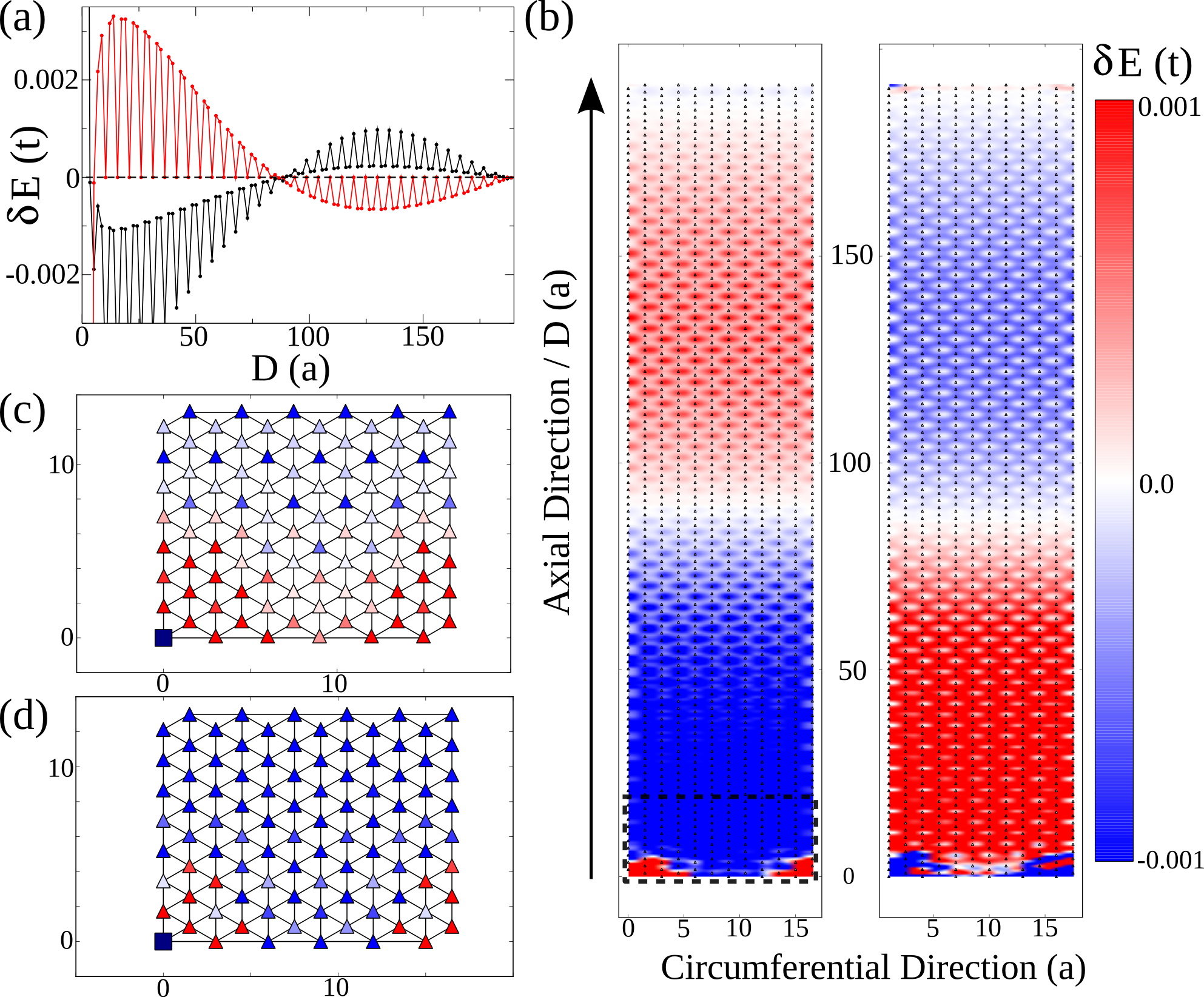}
  \caption{ The interaction $\delta E$ between two hydrogen adatoms, with one fixed to a host at the origin and the second freely moved around to other carbon sites in the system, adsorbed to a ACNT$_6$ host system with $E_F = 0.033 t$.
  Although only two impurities are embedded in the system the $E_F$ value has been shifted to simulate a doped system. 
  The specific value corresponds to an impurity concentration of approximately 1.5\% in the nanotube.
  Oscillations of $\delta E$ with distance along the axial direction are shown in (a) for same sublattice (black) and opposite sublattice (red) configurations.
  The contour plot (b) extends this cross-section to sites in the circumferential direction and the results are shown for same sublattice (left) and opposite sublattice (right) configurations where the short- and long-range behaviour is apparent. 
  For clarity in the illustration, the nanotube has been projected flat and the lattice sites are referred to by their circumferential and axial positions relative to the carbon site hosting the hydrogen impurity at the origin.
  More detailed plots of the short-range behaviour in the left-hand side of (b), i.e. for same sublattice configurations, are shown for the same concentration (c), corresponding to the dashed box in (b), and a lower concentration of 0.1\% (d). }
  \label{fig:deab}
\end{figure}

 The behaviour of $\delta E$ for two hydrogens in a ACNT$_6$ is shown and discussed in Fig. \ref{fig:deab}, the most important characteristics to note are the difference in short- and long-range behaviours. 
 When the impurities are close it is seen that they will prefer to occupy opposite sublattice arrangements, a behaviour occurring in graphene which was noted earlier in this work.
 This short-range behaviour gives way to long range decaying oscillations, complete with circumferential symmetry, and a distinct period 3 behaviour, characteristic of interactions between impurities with zig-zag separation in graphene-like systems, can be observed \cite{gorman2013rkky}.
 The region of short-range behaviour reduces with both increasing concentration, as shown by comparing plots (c) and (d), and increasing CNT circumference.
 The overall short-range and long-range behaviours and the oscillation period are very similar in ZZNTs with the exception that the period 3 behaviour is in the circumferential direction, with long smooth decays along the axial direction. 

\subsection{\label{sec:montecarlo}Monte-Carlo simulation of a finite concentration of hydrogen}

\subsubsection{\label{sec:with_t} Finite Boltzmann temperature ($T_B > 0$)}

\begin{figure}
  \centering
    \includegraphics[width=0.5\textwidth]{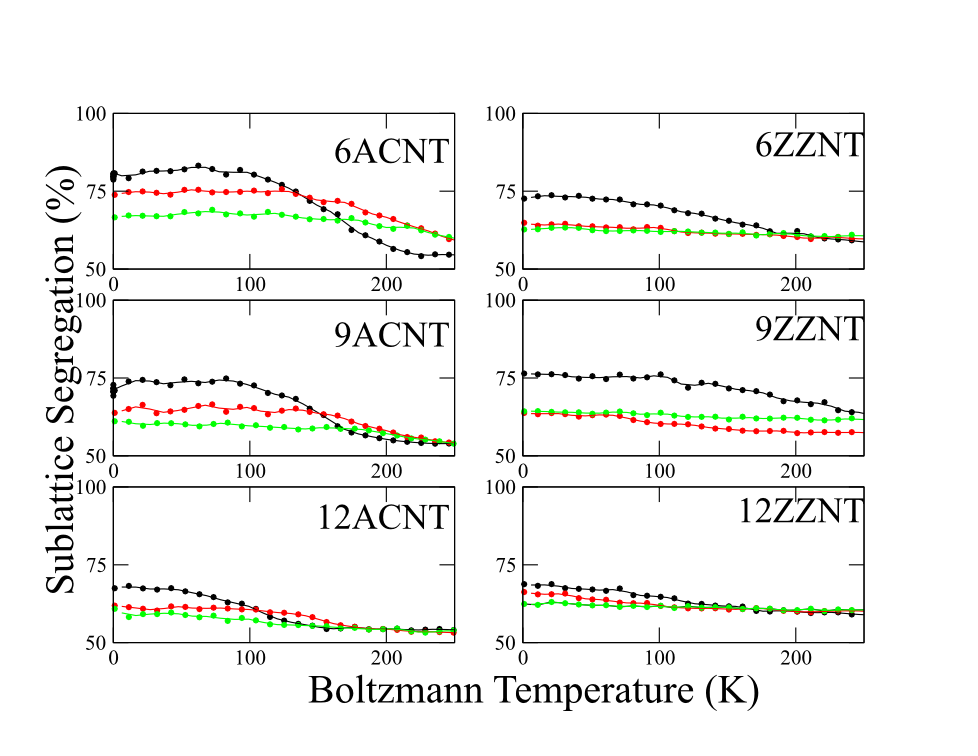}
  \caption{ Degree of sublattice segregation, i.e. percent of impurities on one sublattice, vs. the Boltzmann temperature for small diameter ACNTs and ZZNTs with atomic hydrogen concentrations of 0.5\% (black), 1.0\% (red) and 1.5\% (green). 
  The relationship is analogous to that of magnetism and temperature in the Ising Model.
  It is apparent that increasing concentration leads to a lower expected segregation.
  This arises from the interplay between $E_F$ and impurity separation, both directly linked to the increased concentration of impurities, and this is discussed in detail in Sec. \ref{sec:without_t}.}
  \label{fig:asymm_temp}
\end{figure}

 Monte-Carlo calculations were performed as described in Sec. \ref{sec:mcmethods} to investigate the appearance of sublattice asymmetry in ACNTs and ZZNTs with a range of sizes. 
 The dependence of the sublattice asymmetry on the Boltzmann Temperature for a range of small diameter CNTs with ACNT and ZZNT geometries with different concentrations of hydrogen are shown in Fig. \ref{fig:asymm_temp}. 
 A segregation of 100\% corresponds to all impurities being on one sublattice whilst 50\% corresponds to an equal distribution between the two sublattices.
 As one would expect, increasing $T_B$ causes the impurities to be randomly ordered and the system will approach a 50\% segregation (i.e. unsegregated or symmetric) state. 
 
 The slower decay of the ZZNTs compared to ACNTs to the equilibrium state of 50\% segregation originates from the difference in decay profiles and period 3 behaviour of $\delta E$. 
 In ZZNTs it is harder to push an impurity out of an energetically stable configuration than in an ACNT, hence the broader temperature dependence for the former shown in Fig. \ref{fig:asymm_temp}. The relationship between segregation and Boltzmann Temperature indicates there is no meta-stable state at low temperatures, hence the dependence of segregation on the concentration can be investigated at $T_B=0$ and the result should be a good approximation of the ground state of these systems.
 
\subsubsection{\label{sec:without_t} Zero temperature ($T_B = 0$)}
 
 \begin{figure}
  \centering
    \includegraphics[width=0.40\textwidth]{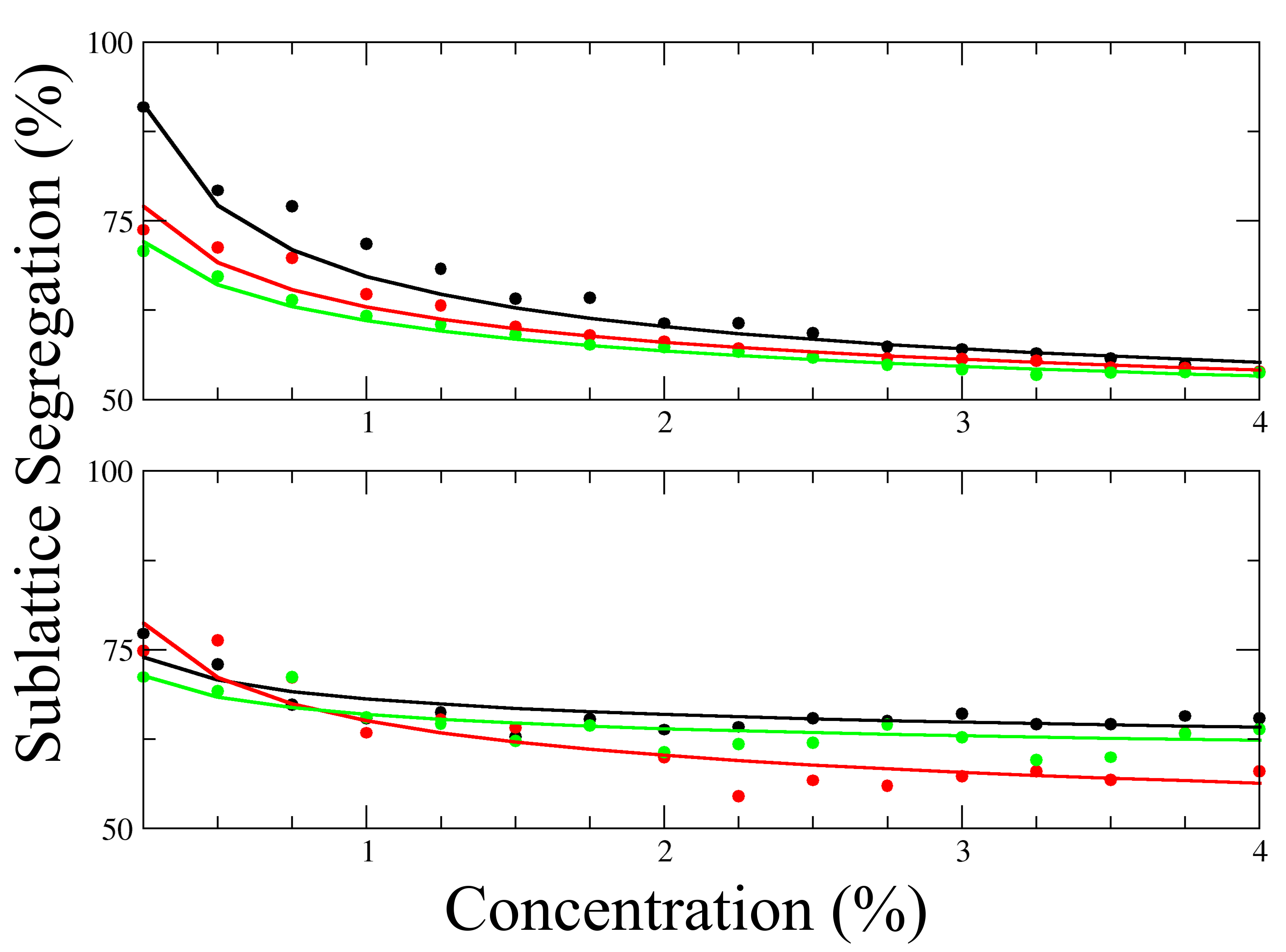}
  \caption{ Segregation vs concentration at $T_B = 0$ for CNT$_6$ (black), CNT$_9$(red) and CNT$_{12}$(green) of armchair (top) and zigzag (bottom) geometry, Monte-Carlo results are represented by the dots and a line of best fit satisfying $\sim 50 + \rho^{-\alpha}$ for the fraction $\alpha \approx \frac{1}{4}$ is shown as a line. Increasing concentration or tube diameter leads to a reduction of the predicted segregation, tending to 50\% in the limit of high concentration, a product of the reduction in the axial separation of impurities leading to a dominance in the short-range behaviour of $\delta E$.}
  \label{fig:asymm_no_temp}
\end{figure}
 
\begin{figure}
  \centering
    \includegraphics[width=0.45\textwidth]{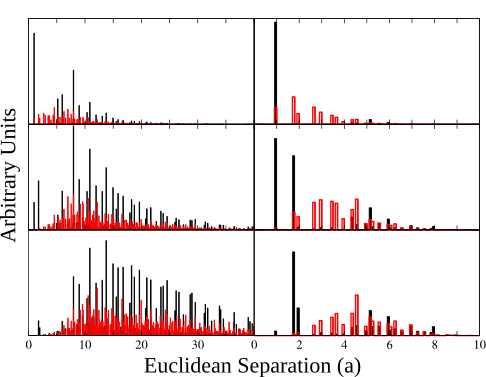}
  \caption{ Histogram data for 1st (top), 2nd (middle) and 3rd (bottom) nearest neighbour separations for H pairs in a ACNT$_6$ at 0.5\% (left) and 5.0\% (right) concentrations, comparing measured data (black) to what one would expect for a random distribution (red).
  The discretised lattice makes identifying the relative sublattice arrangement of the large peaks straightforward.
  Both low and high concentrations show a large peak corresponding to the nearest neighbour opposite sublattice arrangement.
 The data for low concentrations also shows a number of peaks at distances beyond a few lattice parameters, which correspond to same sublattice arrangements in much greater amounts than the random data.
 At high concentrations there are numerous peaks for close separations at much higher frequencies than the random data which points to clustering of the impurities. 
 The 2nd nearest neighbour data shows two close peaks corresponding to opposite (left) and same (right) sublattice arrangements, and these also occur in the 1st and 3rd nearest neighbour data.}
  \label{fig:histograms}
\end{figure}

 The relationship between segregation and concentration is plotted in Fig. \ref{fig:asymm_no_temp} for a range of CNTs and the effect of increasing concentration and CNT diameter can be seen clearly, both leading to a reduction in the predicted segregation effect.
 This can be understood as the long-range $\delta E$ oscillations being the driving mechanism of the asymmetry, as was proposed for the effect for nitrogen impurities in graphene \cite{lawlor2014sublattice}.
 Increasing concentration or the CNT diameter reduces the average axial separation of the impurities and the short-range behaviour, which favours opposite sublattice ordering, dominates.
 This can be corroborated by comparing the k-th nearest neighbour distributions of these systems and this is plotted in Fig. \ref{fig:histograms} comparing low (0.5\%) and high (5.0\%) impurity concentrations, with segregated and unsegregated configurations respectively, and expected values for a random configuration in a ACNT$_6$. 
 Fig. \ref{fig:asymm_no_temp} also suggests the asymmetry is generally expected to be higher in ACNTs than ZZNTs, where levels as high as 90\% segregation should be observable at concentrations around 0.25\% in a ACNT$_6$.

Following the trend of increasing diameter and decreasing segregation it seems a natural conclusion to expand the study to graphene by increasing the nanotube width to infinity, and where naively one should expect no segregation to be observed.
This conclusion should not automatically be made as there are subtle but important distinctions to be made between the case of nanotubes and graphene.
Firstly, pristine graphene is semi-metallic but upon being doped by a finite concentration of hydrogen the band structure is changed considerably, especially around the Dirac point \cite{pike2014tight}, leading to a more complex $E_F$ shift and effects on the LDOS oscillations which drive the segregation effect.
Secondly, these LDOS oscillations behave much differently in graphene decaying as $D^{-2}$ and the directional isotropy of the coupling $\delta E$ would allow for more complete analytic calculations as has been shown for substitutional nitrogen impurities in graphene where a segregation effect was also found \cite{lawlor2014sublattice,lawlor2014sublatticereview,zhao2011visualizing,lv2012nitrogen}.

 Comparison of the nearest neighbour data in Fig. \ref{fig:histograms} demonstrates that clustering occurs in both high and low segregation systems, reflecting previous findings for hydrogen pairs, both in Sec. \ref{sec:twoimps} and in the literature, that when in close proximity they prefer to occupy host sites on opposite sublattices \cite{buchs2007creation,vehvilainen2009multiscale}.
 It has also been shown by Hornaeker et al. that a similar clustering effect can be observed in graphene \cite{hornekaer2006clustering}.
 At lower concentrations (Fig. \ref{fig:histograms} - left) the most frequently found configurations are same sublattice ones, despite the short-range effect, due to the large average axial separation coupled with the long-range oscillatory nature of $\delta E$. 
 This is evident by the location of the peaks in the low concentration data, occurring almost entirely for separations corresponding to same sublattice configurations. 
 Additionally only a small deviation in the expected nearest neighbour profile (with the exception of the short-range peaks) is observed, along with the calculated mean and median values.
 
 These characteristics are in sharp contrast to the high concentration data, shown in Fig. \ref{fig:histograms} (right).
 In this case, there is a much larger degree of clustering, indeed almost all nearest-neighbours will be on the opposite sublattice, and there is a huge difference in the histogram profiles compared to the randomly calculated ones.
 The overwhelming frequency with which one finds opposite sublattice configurations at such short range lends credence to the long-range versus short-range/axial separation hypothesis, where the behaviour of $\delta E$ acts so to cluster the impurities in opposite sublattice configurations at short range.
 
\begin{figure}
  \centering
    \includegraphics[width=0.45\textwidth]{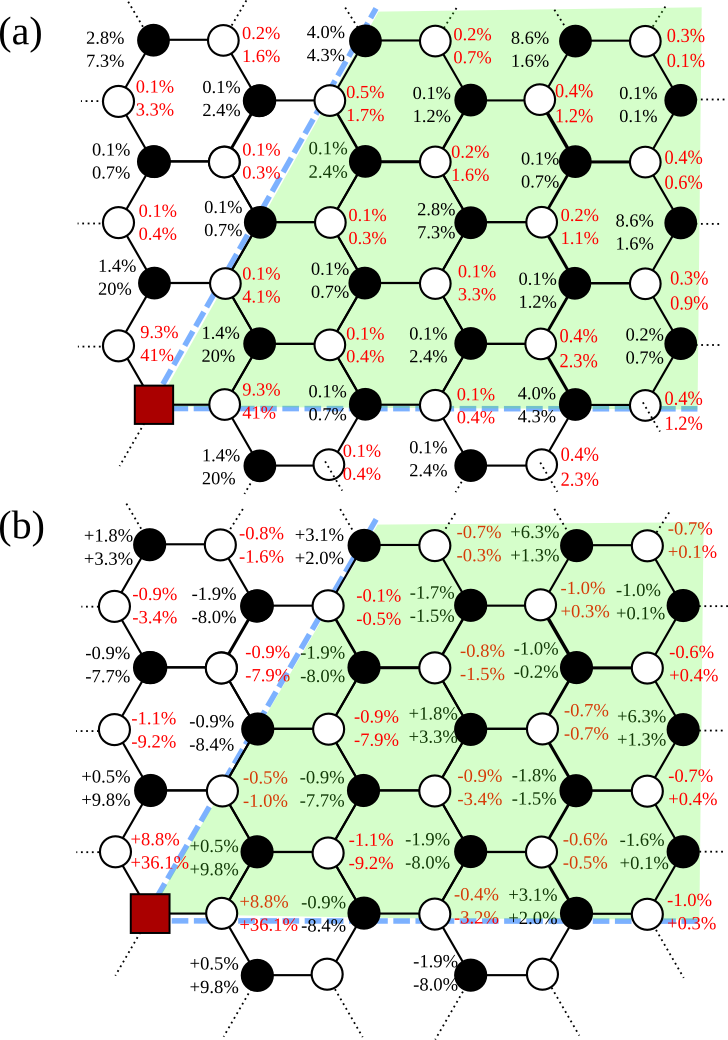}
  \caption{ (a) Approximate probability for an impurity to occupy a host site, given an initial impurity indicated by the red square, for same sublattice (black) and opposite sublattice (red) arrangements at low (top number) and high (bottom number) concentrations in a ACNT$_6$. 
 Due to the symmetry of the system it is possible to allocate irreducible sectors, shown by the green shaded areas, which contain the information on all possible positions in the nanotube in order to simplify the data in this figure.
 To give some context for these findings, part (b) shows the stark difference between the probabilities shown in (a) and values that would be expected for a random, unrelaxed system.}
  \label{fig:grid}
\end{figure}

 A clearer illustration of the clustering effect and the short-range behaviour is shown in Fig. \ref{fig:grid} where the probability of finding impurities in certain short-range configurations is shown and compared for low and high concentrations and the random data, aggregated from the nearest neighbour data of Fig. \ref{fig:histograms}.
 For both high and low concentrations there is a large increase, 8\% and 36\% respectively, in the probability of finding a nearest neighbour same sublattice configuration over what one would expect in a random system. 
 There is also a large increase in same sublattice next-nearest neighbour configurations - a consequence of the clustering of impurities.
 Most interestingly there appear to be preferential configurations, which one would expect from the profile of $\delta E$ in Fig. \ref{fig:deab}.
 For example, there is a 4- or 5-fold increase in likelihood for the same sublattice configuration corresponding to the top-right most black site (position $\approx 7.9a$) and, due to lattice symmetry, equivalent sites such as the middle right-most black site at high and low concentrations. 
 Another preferential configuration is found for a separation of  $\approx 5.2a$ where the probability is approximately doubled.
 This is less pronounced than results seen in hydrogenated graphene \cite{lin2015direct} where much higher concentrations and both sides of the monolayer were used, but is an interesting effect resulting from the inter-impurity interactions.

\section{\label{sec:conclusions}Conclusions}

 In this paper a combination of tight binding Green Functions and Monte Carlo methods have been used to model sublattice segregation of hydrogen adsorbates in carbon nanotubes, motivated by recent findings with sublattice segregation and ordering of impurities in graphene. 
 Understanding this effect is key to tailoring the electronic properties of graphene, nanotubes and perhaps other nanomaterials, along with potential to manufacture of novel adsorbate ordered materials and inducing a magnetic response from graphitic nanosystems.
  
 It was shown in this work that the segregation effect is most pronounced for small diameter nanotubes with dilute concentrations of impurities, and the effect decreases when either of these two parameters are increased.
 The inter-impurity interactions which mediate this interaction have a competing short- and long-range behaviour which tend to prefer opposite and same sublattice configurations respectively, and lead to certain preferential configurations and slight patterning in the lattice. 
 In small diameter tubes with a dilute concentration of impurities the long-range behaviour along the axial direction dominates and the system prefers the segregated state. 
 Increasing the concentration of impurities or the tube diameter reduces this axial separation and the short-range behaviour takes over, forcing the system into a state where the sublattices have approximately equal numbers of dopants.
 The short-range behaviour also leads to clustering and energetically favourable configurations of the impurities, a phenomenon which has been shown before both theoretically and experimentally.

 The current state of research with respect to general hydrogen doping of CNTs \cite{jones1997storage,liu1999hydrogen,orivnakova2011recent,dutta2014review,nikitin2005hydrogenation} and other 1D nanomaterials \cite{bowman2002metallic,gross2002catalyzed,chen2013atomic}, an interest piqued due to their use for hydrogen storage for fuel cells, along with the capability to directly identify dopants and their respective sublattices \cite{tison2013identification, tison2015electronic, joucken2012localized}, suggests that testing the findings of this paper and previous work in graphene \cite{lawlor2014sublattice}, that inter-impurity interactions lead to a segregation effect, is well within current experimental limits.
 
\section{\label{sec:acknowledgements}Acknowledgements }
We acknowledge financial support from the Programme for Research in Third Level Institutions (PRTLI). MSF also acknowledges financial support from Science Foundation Ireland (Grant No. SFI 11/RFP.1/MTR/3083).
We would also like to thank Stephen Power of Technical University of Denmark for useful discussions.

\bibliographystyle{ieeetr}
\bibliography{bibliography.bib}

\end{document}